\documentclass{elsart}

\usepackage{amssymb}

\begin{document}

\begin{frontmatter}

\title{Inhibitory synchrony as a mechanism for attentional gain 
modulation\thanksref{label1}}
\thanks[label1]{To be published in J. Physiol (Paris). Work supported in part by startup
funds from the University of North Carolina and the Sloan-Swartz
Center for Theoretical Neurobiology at the Salk Institute (PT); By
startup funds from Wake Forest University and NINDS grant
NS044894-01 (ES); By the CIRCS fund at Northeastern University (JVJ); 
Also by the Howard Hughes Medical Institute at
the Salk Institute (JMF, TJS).}
\thanks[label8]{Present address JMF: 
Biomedical Engineering department, Duke University, Durham NC 27708-0281}
\author[label1]{Paul H. Tiesinga\corauthref{cor1}}
\ead{tiesinga@physics.unc.edu}
\ead[url]{http://www.physics.unc.edu/~tiesinga}
\corauth[cor1]{Corresponding author,\\
Department of Physics \& Astronomy, Campus box 3255,
University of North Carolina, Chapel Hill, NC 27599-3255;
Fax: 1 919 962 0480; Telephone: 1 919 962 7199}
\author[label3,label4,label8]{Jean-Marc Fellous}
\author[label5]{Emilio Salinas}
\author[label6]{Jorge V. Jos\'e}
\author[label3,label4,label2,label7]{Terrence J. Sejnowski}
\address[label1]{Department of Physics \& Astronomy, 
University of North Carolina, Chapel Hill, NC 27599}
\address[label3]{Computational Neurobiology Lab, Salk Institute,
La Jolla, CA 92037.}
\address[label4]{Howard Hughes Medical Institute, 
Salk Institute, La Jolla, CA 92037.}
\address[label5]{Department of Neurobiology and Anatomy,
Wake Forest University School of Medicine,Winston-Salem, NC 27157}
\address[label6]{Center for the Interdisciplinary Research on Complex Systems
and Physics Department, Northeastern University, Boston, MA 02115.}
\address[label2]{Sloan-Swartz Center for Theoretical Neurobiology, 
Salk Institute, La Jolla, CA 92037.}
\address[label7]{Division of Biological Sciences, University of 
California at San Diego, La Jolla, CA 92093.}

\begin{abstract}
Recordings from area V4 of monkeys have revealed that when the 
focus of attention is on a visual stimulus within the receptive 
field of a cortical neuron, two distinct changes can occur: The 
firing rate of the neuron can change and there can be an increase 
in the coherence between spikes and the local field potential 
(LFP) in the gamma-frequency range (30-50 Hz). The hypothesis 
explored here is that these observed effects of attention could 
be a consequence of changes in the synchrony of local interneuron 
networks. We performed computer simulations of a Hodgkin-Huxley 
type neuron driven by a constant depolarizing current, $I$, 
representing visual stimulation and a modulatory inhibitory input 
representing the effects of attention via local interneuron 
networks.  We observed that the neuron's firing rate and the 
coherence of its output spike train with the synaptic inputs was 
modulated by the degree of synchrony of the inhibitory inputs. 
When inhibitory synchrony increased, the coherence of spiking 
model neurons with the synaptic input increased, but the firing 
rate either increased or remained the same. The mean number of 
synchronous inhibitory inputs was a key determinant of the shape 
of the firing rate versus current ($f-I$) curves. For a large 
number of inhibitory inputs ($\sim 50$), the $f-I$ curve saturated 
for large $I$ and an increase in input synchrony resulted in a 
shift of sensitivity -- the model neuron responded to weaker 
inputs $I$. For a small number ($\sim 10$), the $f-I$ curves were 
nonsaturating and an increase in input synchrony led to an 
increase in the gain of the response -- the firing rate in 
response to the same input was multiplied by an approximately 
constant factor. The firing rate modulation with inhibitory 
synchrony was highest when the input network oscillated in the 
gamma frequency range. Thus, the observed changes in firing rate 
and coherence of neurons in the visual cortex could be controlled 
by top-down inputs that regulated the coherence in the activity 
of a local inhibitory network discharging at gamma frequencies.
\end{abstract}

\end{frontmatter}
\setlength{\parindent}{0.2in}
\section{Introduction}

Only a small part of the information in natural visual scenes is 
consciously accessible following a brief presentation 
\cite{Simons2000,Rensink2000}. The information that is retained 
depends on what is attended \cite{CognNeuroBook}. The neural 
correlates of selective attention have been studied in monkeys 
using recordings from single neurons 
\cite{Connor96,Connor97,Luck1997,McAdams1999,Treue1999,Reynolds2000,Fries2001,Treue2002,Moore2003,Reynolds2003}. 
A key finding is that attention modulates both the mean firing 
rate of a neuron in response to a stimulus 
\cite{McAdams1999,Reynolds2000,Reynolds2003} and the coherence of 
spikes with other neurons responsive to the stimulus 
\cite{Steinmetz2000,Fries2001}. The increase of coherence with 
attention is strongest in the gamma frequency range 
\cite{Fries2003}.\\
\indent 
There are many different types of inhibitory interneurons in the 
cortex, each with different patterns of input and output.  For 
example, the basket cells project almost exclusively onto the 
somas and proximal dendrites of pyramidal neurons.  In addition 
to their function in suppressing cortical activity, inhibitory 
cells may also be responsible for shaping the temporal pattern of 
spiking activity in the cortical network.  In particular, the 
synapses from baskets cells near the spike initiating zone are 
effective in gating the occurrence of spikes \cite{Lytton91,Cobb95}, 
and since a single basket cell has multiple inhibitory 
contacts on several thousand pyramidal cells, it could 
synchronize a subset of active cortical neurons.\\
\indent 
Local cortical interneurons are not isolated from each other but 
form networks connected by fast GABAergic inhibitory synapses and 
electrical gap junctions 
\cite{Galarreta1999,Gibson1999,Beierlein2000,Galarreta2001a,Galarreta2001}. Networks of inhibitory interneurons have been 
implicated in the generation of synchronous gamma-frequency-range 
oscillations in the hippocampus 
\cite{whittington,traub,Buhl98,Hormuzdi2001} and the cortex 
\cite{Deans2001}, and could entrain a large number of principal 
cells \cite{BushSejn96,Koos99,Tamas00,Galarreta2001} as well as 
modulate their firing rates \cite{Soltesz2002a}.  Hence, 
interneuron networks could mediate the effects of attention 
observed in cortical neurons.\\
\indent 
We recently proposed a mechanism, {\em synchrony by competition}, 
for rapid synchrony modulation in interneuron networks 
\cite{RapidMod}(see Discussion), that could explain how attention 
modulates the synchrony of interneuron networks.  The focus of 
this study is on the impact of synchronous inhibitory input onto 
the principal output neurons of the cortex. \\
\indent 
The degree of synchrony of inhibitory inputs can be characterized 
by their temporal dispersion, referred to here as jitter.   A 
lower value of jitter corresponds to more synchrony and means 
that inhibitory inputs tend to arrive at the same time.  We 
explore how the response of the output neuron -- the firing rate 
and coherence with inhibitory oscillation -- is modulated by 
jitter. Our main results are: (1) The mean number of inhibitory 
inputs determines whether variations in synchrony lead to changes 
in the neuron's sensitivity or whether they modulate the gain of 
the response (Section 3.2, 3.3). (2) The response modulation by 
synchrony is most potent at gamma frequencies (Section 3.4). (3) 
Modulation of the single neuron response properties by inhibitory 
synchrony is consistent with the effects of attention observed 
{\em in vivo} \cite{McAdams1999,Fries2001} (Section 3.5). See the 
Appendix for a review of previous experimental results. 
Earlier reports of these results have appeared as abstracts 
\cite{GainSFN,GainModSFN02}.

\section{Methods}
\subsection{Modeling synchronous inhibitory inputs}
\indent 
In a previously studied model of inhibitory interneurons connected 
by chemical synapses, the network produced oscillatory activity that 
consisted of a sequence of synchronized spike volleys 
\cite{MutInhibPub,Aertsen99}. 
First, we describe the statistics of the output of the
interneuron network. Each spike volley was characterized by the
number of spikes in the volley $a_{IV}^i$ (with $i$ the volley
index, $a$ the activity and IV indicating inhibitory volley),
their mean spike time  $t_{IV}^i$ and their spike-time dispersion
$\sigma_{IV}^i$. There was variability in the volleys:
\begin{itemize}
\item{}The number of spikes varied across cycles. 
We used Poisson statistics, hence, the mean of $a_{IV}^i$ 
across cycles was $a_{IV}$ and its
standard deviation was $\sqrt{a_{IV}}$.
\item{} The network oscillations were not perfectly regular. There was
a stochastic drift in the interval between the arrival of consecutive volleys.
The interval, $P_i\equiv t_{IV}^{i+1}-t_{IV}^i$, was approximately normally
distributed with mean (the period) $P$ and
a coefficient of variation equal to $CV_T$ (we use
$CV_T$ rather than $CV_P$ to avoid confusion with a previously
introduced synchrony measure \cite{RapidMod}). 
The oscillation frequency was $f_{osc}=1/P$.
\item{} The mean of $\sigma_{IV}^i$ across cycles was $\sigma_{IV}$.
\end{itemize}
\indent 
The spike-time dispersion, $\sigma_{IV}$, is inversely related to 
the synchrony of the network oscillation.  A synchronized network 
has $\sigma_{IV}=1$ ms ($1~ms$ is the order of magnitude of the 
jitter caused by intrinsic noise in cortical slices 
\cite{Mainen95}), whereas, for gamma oscillations, 
$\sigma_{IV}=10$ ms corresponds to an asynchronous network 
\cite{MutInhibPub}.  Hence, in the simulations of the model, 
$\sigma_{IV}$ was varied between $1$ and $10$ ms. The mean number 
of spikes per volley, $a_{IV}$, is determined by the fraction of 
network neurons that is active on a given cycle, the size of the 
network, and the cortical GABAergic presynaptic release 
probability \cite{KochBook}. The network of interneurons was not 
explicitly simulated; rather the above statistics were used to 
model input spike trains representing the synchronous inhibitory 
input, as described below. These input spike trains will be 
referred to as `network activity' throughout.\\
\indent 
The method used to obtain synchronous volleys is illustrated in 
Fig.$~$\ref{Methods}. First, a set of volley times $t_{IV}^i$ was 
generated (with mean intervolley interval $P$ and a coefficient 
of variation $CV_T$). Next, a binned spike-time probability (STP) 
was obtained by convolving the volley times with a Gaussian 
filter with standard deviation $\sigma_{IV}$ and area $a_{IV} \Delta t$ (the 
bin width $\Delta t=0.01$ ms was equal to the integration time-step 
used in the simulations). The Gaussian filter was 40 ms long
in order to accomodate at least 2 standard deviations for
the maximum $\sigma_{IV}$ used in the simulations.  The peak of the 
Gaussian was located at the center of the filter at 20 ms. As a result, the  
effect of changing $\sigma_{IV}$ during the simulation acts 
with a delay of $20$ ms. Input spike times were generated as a 
Poisson process from the STP, as in \cite{NetworkOptim}. Each 
input spike produced an exponentially decaying conductance pulse, 
$\Delta g_{inh} \exp(-t/\tau_{inh})$ in the postsynaptic cell 
($inh=$inhibitory), yielding a current $I_{syn}=\Delta g_{inh} 
\exp(-t/\tau_{inh}) (V-E_{GABA})$. In this expression $t$ is the 
time since the last presynaptic spike, $\tau_{inh}$ is a decay 
time constant, $\Delta g_{inh}$ is the unitary synaptic 
conductance, $V$ is the postsynaptic membrane potential, and 
$E_{GABA}=-75~mV$, is the reversal potential. The values of these 
parameters were varied and their specific values are given in 
each figure caption (see Appendix). A similar procedure was used for 
synchronous excitatory inputs. The same notation holds with $EV$ 
(excitatory volleys) replacing $IV$ and $exc$ (excitatory) 
replacing $inh$.\\
\indent
The resulting train of conductance pulses drove a single 
compartment neuron with Hodgkin-Huxley voltage-gated sodium and 
potassium channels, a passive leak current, synaptic currents as 
described above and a white noise current with mean $I$ and 
variance $2D$. Full model equations are given in the Appendix 
\cite{MutInhibPub}. They were integrated using a noise-adapted 
2nd-order Runge-Kutta method \cite{greenside}, with time step 
$dt=$0.01 ms. The accuracy of this integration method was checked 
for the dynamical equations without noise ($D=0$) by varying $dt$ 
and comparing the result to the one obtained with the standard 
4th-order Runge-Kutta method \cite{Press} with a time-step $dt$ of 
$0.05~ms$.

\subsection{Statistical analysis}
\indent 
Simulations were run multiple times with different seeds for the 
random number generator, yielding different trials.  Spike times 
$t^j_i$ ($i$th spike time during the $j$th trial) of the target 
neuron were calculated as the time that the membrane potential 
crossed $0$ mV from below. The mean interspike interval, $\tau$, 
was calculated as the mean of all intervals during a given trial 
and then averaged across all trials. The mean firing rate $f$ 
was $1000/\tau$ ($\tau$ is in ms, $f$ in Hz).  The coefficient 
of variation (CV) was the standard deviation of the interspike 
intervals across one trial divided by their mean. The CV was then
averaged across all trials. Errors in these statistics were 
estimated as the standard deviation across $10$ equal-sized 
subsets of the data. The Fano factor (FF) was the variance of 
the spike count in a given time interval divided by the mean 
spike count.\\
\indent
For neurons receiving a synchronous inhibitory input,
the firing rate versus current ($f-I$) curves were obtained
by systematically varying the depolarizing input current $I$ and 
calculating the firing rate as described above.
$f-I$ curves were fitted to a sigmoidal function,
$f(I)=\frac{A}{2}(1+\tanh(\lambda_I (I-\Delta_I))$.
We also attempted
to make the curves for different values of $\sigma_{IV}$
overlap with a reference curve using the following three procedures:
(1) A shift of $I$ over $\Delta_I$;
(2) A multiplication of $I$ by $\lambda_I$;
(3) A multiplication of $f$ by $\lambda_f$.
The fitting procedures involved the MATLAB routine ${\tt nonlinfit}$.
Confidence intervals at 95\% for the fitting parameters were obtained
using the routine ${\tt nlparci}$.
Four different fits were used to minimize the difference
between the fitted curve and the reference curve:
(1) Shift of $I$; 
(2) Shift and scaling of $I$; 
(3) Shift of $I$ and scaling of $f$; 
(4) Shift of $I$ and scaling of $I$ and $f$. 
In most cases fit (3) yielded the best
results or yielded results that were close to the best 
three-parameter fit (4).
For the purpose of comparison we show only the results of fit (3).\\
\indent
The coherence of the output spike times with the underlying 
oscillations was determined using the spike-triggered average 
(STA) and the vector strength (VS). The local field potential 
(LFP) was estimated as the membrane potential of a model neuron 
receiving coherent synaptic inputs while being hyperpolarized to 
prevent action potentials. Spike times were obtained from 
another neuron receiving the same synaptic inputs. The STA was 
then calculated by taking the membrane potential from the first 
(hyperpolarized) neuron centered on the spike times of the 
second (depolarized) neuron \cite{AbbottBook}.  The STA power 
spectrum was calculated using the MATLAB routine {\tt psd} with 
standard windowing using a $2048$ point fast Fourier transform 
(sampled at $0.2$ ms). The spike-field coherence (SFC) was 
computed as the STA power spectrum divided by the power spectrum 
of the LFP, as was done in the analysis of the  experimental 
data \cite{Fries2001}. \\
\indent
The spike-phase coherence was determined from the variance and
mean value of the phase of the spike time $t$ with respect to the
oscillations. The phase was defined as
$\phi=(t-t_{IV}^i)/(t_{IV}^{i+1}-t_{IV}^i)$ \cite{Moss99}. Here
$t_{IV}^i$ is the last volley time before $t$ and  $t_{IV}^{i+1}$
is the first volley time after $t$. We determined the standard
deviation $\sigma_{\phi}$ of $\phi$ and the VS \cite{Mardia2000},
\begin{equation}
VS=\sqrt{\langle \cos(2\pi \phi) \rangle^2 +\langle \sin(2\pi \phi)\rangle^2}.
\end{equation}
Here $\langle \cdot \rangle$ is the average over all
phases in a given trial and across all trials. 
VS is zero when $\phi$ is uniformly
distributed between zero and one, and is one when the phase is constant.

\section{Results}
\subsection{Inhibitory input synchrony modulates output firing 
rate and coherence}
\indent 
The goal is to reproduce the effects of attention on cortical 
neurons in the visual cortex.  In the model neuron that we 
simulated (Fig.$~$\ref{ExampleSpikeTrainRegular}) the stimulus-induced 
activation was represented by injecting an
excitatory input current. In this baseline state, the object in 
the receptive field is not being attended. We explore the 
hypothesis that when attention is directed to an object in the 
neuron's receptive field, the discharge of the neuron is 
modulated through inhibitory inputs.\\
\indent
Fig.$~$\ref{ExampleSpikeTrainRegular}A shows the voltage 
response of a model neuron driven by an inhibitory input 
oscillating at approximately 40 Hz. On each cycle of the 
oscillatory drive it received a volley with an average of 
$a_{IV}=25$ pulses, each with a unitary peak conductance of 
$\Delta g_{inh}=0.044 mS/cm^2$. The time-average of the 
inhibitory conductance was $0.44 mS/cm^2$, which was about four 
times larger than the leak conductance $g_L=0.1 mS/cm^2$. The 
neuron was not spontaneously active in the absence of synaptic 
inputs; hence, in order to make it spike in the presence of 
inhibition, a constant depolarizing current $I=4.0 \mu A/cm^2$ 
was also injected. The temporal dispersion of the input spike 
times was on average $\sigma_{IV}$. In the baseline state, 
$\sigma_{IV}=8~ms$, the inhibitory input was asynchronous, but 
during the time interval $t=300~ms$ and $700~ms$ the input was 
made synchronous by decreasing $\sigma_{IV}$ to $2~ms$. During 
the baseline state the neuron fired at a low average rate 
($f=4.4\pm 0.7~ Hz$ obtained over a longer segment, 
$1000~ms$, than shown in the figure). The spiking statistics for 
this case and the figures below are summarized in Table 1. \\
\indent
When the inhibitory input was synchronous, the firing rate 
increased by a factor of four to $f=18.3\pm 0.4~ Hz$. This 
increase was robust across trials as indicated by the spike time 
histogram across $500$ trials 
(Fig.$~$\ref{ExampleSpikeTrainRegular}C) and the rastergram for 
the first ten trials (Fig.$~$\ref{ExampleSpikeTrainRegular}D). 
Note that the time-averaged inhibitory conductance remained 
constant during the entire trial, even during the episode of 
enhanced input synchrony. Thus, the increase in firing rate was 
due solely to the change in coherence.\\ 
\indent
In Fig.$~$\ref{ExampleSpikeTrainRegular}, a neuron was activated 
by a suitable stimulus that is ignored: initially, the neuron 
did not respond and the presence of the stimulus was not 
transmitted to downstream cortical areas, but during the time 
interval of increased input synchrony, the presence of the 
stimulus was signaled to downstream areas.  The synchrony of the 
inhibitory input acted as gate. For the parameters of 
Fig.$~$\ref{ExampleSpikeTrainRegular}, when the stimulus was 
presented for a short interval, 400 ms, the neuron did not 
produce a spike on most trials during the baseline state, but it 
did produce a couple of spikes during the period of increased 
input synchrony. The next three Sections focus on the 
statistical properties of the neuron's output spike train -- how 
the firing rate and the coherence with the synaptic inputs 
varied with the parameters of the inhibitory input, 
specifically, $a_{IV}$, $\sigma_{IV}$ and the oscillation 
frequency $f_{osc}$.\\
\indent
The spike trains of cortical neurons recorded in vivo are highly 
variable in time and across trials \cite{Shadlen98}. The spike 
trains obtained on different trials (see the rastergram in 
Fig.$~$\ref{ExampleSpikeTrainRegular}D) were significantly 
different from each other and the interspike intervals ranged 
from $5~$ms to the $100s$ of ms. The irregularity of spike 
trains in time was quantified by the coefficient of variation 
(CV): the standard deviation of the interspike intervals {\em in 
time}, divided by their mean. CV values between $0.5$ and $1.0$ 
are typical for {\em in vivo} recordings \cite{Shadlen98}. The 
values obtained for Fig.$~$\ref{ExampleSpikeTrainRegular} were 
in this range, CV$=0.96\pm 0.14$ during the baseline state and 
CV$=0.83\pm 0.03$ during the interval with increased input 
synchrony. We used longer segments than shown in 
Fig.$~$\ref{ExampleSpikeTrainRegular} -- 1000 ms across 500 
trials -- to obtain more robust estimates for the CV values. 
In Fig.$~$\ref{ExampleSpikeTrainRegular}A the spike 
train looked quite regular during the period with high input 
synchrony because it was entrained to the input oscillation. 
Although the interspike intervals were variable they were 
approximate multiples of the oscillation period. Hence, the high 
CV value was a consequence of the multimodality of the 
interspike interval distribution.\\
\indent 
The irregularity of spike trains across trials is commonly 
expressed as the Fano factor (FF): the variance of the spike 
counts during a time interval divided by the mean. For a 
$1000~ms$ interval, FF$=1.2 \pm 0.2$ (baseline) and FF$=0.67\pm 
0.09$ (increased input synchrony). In comparison, a homogeneous 
Poisson process has CV$=1$ and FF$=1$. The value of the Fano 
factor is a direct consequence of our choice of model 
parameters: there is variability across trials in the arrival 
time of the inhibitory volley, $CV_T$, variability in the number 
and the timing of the inputs in the volley, and intrinsic noise. 
Consistent with this expectation, much smaller Fano factors were 
obtained when one or more of these sources of variability were 
absent. \\
\indent
The spike-field coherence (SFC) has been used to quantify the 
degree of coherence of spike trains with the local field 
potential (LFP). The LFP is measured using an extracellular 
electrode and is assumed to reflect the synaptic inputs to 
neurons close to the electrode and their resulting activity 
\cite{Frost67}. It is not known in general how many neurons 
contribute to the LFP and whether it is dominated by the 
activity of local neurons or whether it more closely reflects 
the synaptic activity due to presynaptic neurons. We calculated 
the LFP as the membrane potential fluctuations in a neuron that 
was hyperpolarized to prevent action potentials. The SFC in the 
gamma frequency range (34-44 Hz) decreased from $0.14\pm0.14$ in 
the baseline state to $0.026\pm 0.010$ during the period of 
increased inhibitory input synchrony. This counterintuitive 
result underscores the fact that in modeling studies the SFC so 
defined may not be the best way of calculating the coherence; 
hence, the vector strength (VS) will be used here instead, 
although for comparison with experiments in Section 3.5 the SFC 
will be used. The relation between the SFC and the VS is explained in 
the Appendix. During the state of increased inhibitory input 
synchrony, the VS increased to $0.878\pm 0.006$ from $0.710\pm 
0.045$ in the baseline state.  The key constraint is that one 
needs enough spikes to accurately estimate the SFC. The SFC for 
the baseline state was recalculated, but with the injected 
current increased from $4.0 \mu A/cm^2$ to $6.0\mu A/cm^2$ in 
order to make the neuron's firing rate similar to that for 
synchronous inputs. The SFC was $0.008\pm0.007$ in the gamma 
range and $0.016\pm0.012$ in the theta range. \\
\indent
Fig.$~$\ref{ExampleSpikeTrain} shows another example of firing 
rate modulation with input synchrony. The model neuron was again 
driven by a synchronous inhibitory drive, but part of the 
depolarization was provided by a homogeneous excitatory Poisson 
process. 
During the 
time interval between $t=1000$ and $2000$ ms, $\sigma_{IV}$ was 
decreased from $4$ ms to $2$ ms. For $t<1000$ ms, the firing 
rate was $22.3$ Hz and $VS=0.685\pm 0.012$ (other statistics are 
listed in Table 1). When the input synchrony was increased the 
firing rate increased to $34.6~$Hz and $VS=0.744\pm 0.004$. In 
contrast to Fig.$~$\ref{ExampleSpikeTrainRegular}, the spike 
trains looked more like the ones found {\em in vivo}. The degree 
of synchrony also modulated the firing rate at higher values 
(Fig.$~$\ref{ExampleSpikeTrain}), rather than just acting as a 
gate (Fig.$~$\ref{ExampleSpikeTrainRegular}). \\ 
\indent In both examples the model predicts that 
synchrony-induced increases in firing rate are associated with a 
decrease in firing variability and an increase in coherence.

\subsection{Modulation of $f-I$ curves with input synchrony}
\indent 
The firing rate was plotted as a function of the depolarizing 
current $I$ and the jitter $\sigma_{IV}$ in  
(Fig.$~$\ref{FirSurf1}A).  The firing rate ranged from $0$ to 
$80~$Hz for $1~ms\le \sigma_{IV} \le 6~ ms$ and $2~\mu A/cm^2 
\le I \le 7\mu A/cm^2$. An increase in $\sigma_{IV}$ usually led 
to a reduction in firing rate (dashed arrow in 
Fig.$~$\ref{FirSurf1}A). The $f-I$ curve for constant 
$\sigma_{IV}$ had a knee for small values of $\sigma_{IV}$ 
(arrow in Fig.$~$\ref{FirSurf1}A). At this point in the neuron's 
operating range, decreasing $\sigma_{IV}$ (increasing input 
synchrony) did not lead to an increase in firing rate, but the 
coherence did increase (see below). Neither did an increase in 
driving current result in a higher firing rate. Because of the 
stochastic nature of the dynamics the firing rate could be 
arbitrarily low. For practical purposes the onset of firing was 
defined as the point where the firing rate exceeded $1~Hz$. This 
means that on average more than one spike should be observed on 
$1$~s long trials. The current value at which this happened 
increased steadily with increasing $\sigma_{IV}$ 
(Fig.$~$\ref{FirSurf1}A).\\
\indent
To determine how the variability and the coherence were related 
to the firing rate $f$, CV-f and VS-f plots were made at 
constant $\sigma_{IV}$ values (Fig.$~$\ref{FirSurf1}B,C). Each 
data point represented a point on the $f$ versus $I$ and 
$\sigma_{IV}$ surface. The CV was $1$ for small firing rates, 
but decreased with $f$ until $f\approx 40~Hz$, at which point 
the CV increased with $f$ again. The minimum CV value reached at 
$f\approx 40 ~Hz$ was $CV\approx 0.1$ for $\sigma_{IV}=1~ms$ and 
increased with $\sigma_{IV}$.  The relationship between the knee
in the $f$-$I$ curve and the dip in the $CV$-$f$ curve can be
understood in terms of phase locking to the synchronous inhibitory
synaptic drive. When the neuron is phase-locked it fires one 
spike on each cycle and the firing rate is 40 Hz. The interspike 
interval is approximately equal to duration of an oscillation
cycle, but there is some variability due to jitter of the spike time 
within each cycle. The $CV$ is small for this situation. 
For smaller driving currents,
the phase locking becomes less stable and the firing rate falls 
below 40Hz.
The neuron will then skip cycles leading to  a bimodal 
distribution of interspike intervals: the intervals are either 
approximately equal to one 
cycle or two cycles. As a result the $CV$ increases sharply. For larger 
driving currents phase locking also becomes unstable, but now the firing 
rate exceeds 40Hz. 
On some cycles there are more than one spike, 
yielding a bimodal interspike interval distribution. The 
 CV also increases sharply for this situation. A more detailed 
description of 
phase locking to a periodic inhibitory drive can be found in 
Ref.\cite{NetworkOptim}.\\
\indent
The most regular spike trains were 
obtained near the knee in the firing rate surface.  The VS had 
its highest value for small firing rates and decreased very 
slowly with firing rate up to $f\approx 40~Hz$. For $f>40~Hz$, 
there was a precipitous drop in VS with firing rate. This 
occurred because for $f<40~Hz$ the neuron produced at most one 
spike per cycle at approximately the same phase with respect to 
the oscillation -- hence the VS was high, but for $f>40~Hz$, 
there could be two spikes on a given cycle, at two different 
phases, and hence the VS was lower. The behavior of the FF
was similar to that of the CV (Fig.$~$\ref{FirSurf1}D).\\
\indent
We performed the same analysis on a model neuron with the 
parameters used to generate the spike trains in 
Fig.$~$\ref{ExampleSpikeTrain}. The firing rate surface is shown 
in Fig.$~$\ref{FirSurf2}A. There was no `knee'; instead, the 
firing rate always decreased with increasing $\sigma_{IV}$ and 
always increased with $I$. The variability of the spike trains 
(Fig.$~$\ref{FirSurf2}B) was highest for low firing 
rates and the CV took values ranging between $CV=0.9$ and $1.3$. 
The CV remained constant for the highest $\sigma_{IV}=7~ms$ to 
$10~ms$ values studied, but decreased with firing rate for lower 
$\sigma_{IV}$ values. There was no evidence for a minimum at 
$f\approx 40~Hz$, in contrast to the preceding case. The VS 
decreased linearly as a function of firing rate 
(Fig.$~$\ref{FirSurf2}C) and decreased in all cases 
with $\sigma_{IV}$. The behavior of the FF
was similar to that of the CV (Fig.$~$\ref{FirSurf2}D).\\
\indent
The simulations for which part of the depolarizing drive is provided by 
excitatory synaptic inputs are more realistic.
However, when the rate of the excitatory inputs is varied, both the
mean as well as the variance of the current fluctuations are altered
(see, for instance, Ref.\cite{TiesPRE}). This makes it harder
to distinguish changes due to the mean driving current from those
due to the variance in the driving current. 
For this reason, the following analysis is only performed on
neurons driven by a depolarizing current.\\
\indent
The shape of the $f-I$ curves varied with 
$\sigma_{IV}$ for $a_{IV}=10$ and $a_{IV}=50$ (Figure 
\ref{fICurves}A,B). For $a_{IV}=10$, the $f-I$ curves were 
nonsaturating (Figure \ref{fICurves}A), but could not be fitted 
by a power law relationship between $f$ and $I$. In previous 
studies it was found that the sensitivity and the gain of $f-I$ 
curves were modulated by the statistics of the synaptic input 
\cite{Silver2003,JM_GAIN}. Therefore we tried to collapse the $f-I$ 
curves for different $\sigma_{IV}$ onto a reference curve, here 
taken to be $\sigma_{IV}=1~ms$, by rescaling the firing rate 
(gain change: $f\rightarrow f/\lambda_f$) and changing the 
sensitivity (shift in $I$, $I\rightarrow I-\Delta_I$). The 
fitting parameters, $\lambda_f$ and $\Delta_I$, were plotted 
as a function of $\sigma_{IV}$ in Figure \ref{fICurves}C 
(for $\sigma_{IV}=1~ms$, by 
definition, $\lambda_f=1$ and $\Delta_I=0$). For  $a_{IV}=50$, 
the $f-I$ curves saturated at approximately $40~Hz$ (Figure 
\ref{fICurves}B), which corresponded to the knee in 
Fig.$~$\ref{FirSurf1}. It was not possible to scale 
the curves by $\lambda_f$ in order to make them collapse since 
that would alter the saturation value.
For $\sigma_{IV}\le 4~ms$, the $f-I$ curves were well fitted by 
a sigmoid function, $f=A/2(1+\tanh(\lambda_I (I-\Delta_I))$. 
The fitting parameter 
$\Delta_I$ corresponded again to a shift in sensitivity, 
$\lambda_I$ corresponded to the slope of the sigmoid and $A$ was 
the saturation value of the firing rate.  The higher the value 
of $\lambda_I$ the smaller the range of current values over 
which the firing rate went from firing rates near zero to 
$f\approx A$. For firing rates $f\ll A/2$, this can be 
interpreted as a change in gain, with the gain proportional to 
$\lambda_I$. We performed the fitting procedure with $A$ as a 
free parameter, but this did not result in a better fit compared 
with the fit obtained by setting $A$ equal to $f_{osc}$. The 
fitting parameters $\Delta_I$ and $\lambda_I$ are plotted versus 
$\sigma_{IV}$ in Fig.$~$\ref{fICurves}D. Increasing 
$\sigma_{IV}$ led to a shift to the right ($\Delta_I>0$), 
decreasing sensitivity and stretching the sigmoid ($\lambda_I$ 
decreased), yielding a larger dynamical range.\\
\indent
The shift in sensitivity going from $\sigma_{IV}=5~ms$ to 
$\sigma_{IV}=1~ms$ was much higher for $a_{IV}=50$ than for 
$a_{IV}=10$. In contrast, for $a_{IV}=10$, the gain change was 
more pronounced than for $a_{IV}=50$ and it also extended over a 
larger range.  The difference between $a_{IV}=10$ and $a_{IV}=50$ 
persisted when
part of the depolarizing current was provided by excitatory synaptic
inputs. We determined the $f$-$I$ curves for $a_{IV}=10$ (Fig.$~$\ref{fICurvesExtra}A)
and $a_{IV}=50$ (Fig.$~$\ref{fICurvesExtra}B) while the neuron was 
driven by varying amounts of excitatory synaptic inputs 
($\sigma_{IV}$ was kept
fixed at 1 ms). There was no knee in the $f$-$I$ curve 
for $a_{IV}=10$. In contrast, there was a knee for $a_{IV}=50$
that persisted for input rates up 700 EPSPs per second. 
The knee was accompanied by a dip in the 
CV versus $I$ curve (Fig.$~$\ref{fICurvesExtra}D). For $a_{IV}=10$
(Fig.$~$\ref{fICurvesExtra}C),
the dip was much less pronounced or absent.\\
\indent
In summary, for small $a_{IV}$, changes in gain with input 
synchrony dominated, whereas for large $a_{IV}$, changes in 
sensitivity dominated. The mean number of inputs on each 
oscillation cycle, $a_{IV}$, was an important characteristic of 
the presynaptic network and could be modulated (see Discussion).

\subsection{Shift in sensitivity versus change of gain}
\indent 
The responses of neurons driven by excitatory and inhibitory 
synaptic inputs have been studied extensively (see Discussion). 
A change in the mean conductance of the synaptic inputs results 
in a shift of the $f-I$ curve, whereas a change in the variance 
of the conductance corresponds to a change in gain 
\cite{ChanceGain,JM_GAIN}. Is it possible to understand the 
results of the preceding section in terms of the `mean' and the 
`variance'?\\ 
\indent 
The mean and variance of the conductance can be calculated 
either by taking one long trial and performing a time average or 
averaging an observation of the variable at one specific time 
across many trials. For a stationary process, for which the 
statistics do not change in time, these two methods are 
equivalent. The time-averaged conductance of the synchronous 
inhibitory input does not depend on the jitter, $\sigma_{IV}$, 
but the neuron's response does. Hence, one would infer that the 
change in firing rate with $\sigma_{IV}$ would be due solely to 
the `variance'.\\ 
\indent 
In Fig.$~$\ref{InhibCond}Aa, dashed line, we show the inhibitory 
conductance as a function of time on a given trial. There was a 
peak every $25~ms$, but its amplitude was different on each 
cycle because the number of inputs and their timing differed 
across cycles. However, the conductance waveform averaged across 
trials (Fig.$~$\ref{InhibCond}Aa, solid line) was the same on 
each cycle. Furthermore, the conductance waveform obtained on 
one cycle by averaging across all cycles in one trial 
(Fig.$~$\ref{InhibCond}Ab), was the same as the trial average. 
The variability across trials was also the same as the 
variability across cycles and is visualized in 
Fig.$~$\ref{InhibCond}Ab as the mean plus or minus twice the 
standard deviation. In the following we refer to either the mean 
or variability without specifying whether it is across trials or 
cycles.  We also kept the time-averaged mean conductance 
constant while varying $a_{IV}$ by scaling the unitary 
conductance as $1/a_{IV}$. The mean conductance waveform did not 
depend on $a_{IV}$, but the variability decreased with 
increasing $a_{IV}$ (Fig.$~$\ref{InhibCond}Ba). The mean 
conductance waveform as well as the variability depended on 
$\sigma_{IV}$ (Fig.$~$\ref{InhibCond}Bb).\\
\indent
When a neuron is driven by a time-varying inhibitory conductance 
and the injected current is in an appropriate range, it can only 
spike when the inhibitory conductance is small. There is a value 
for the inhibitory conductance above which the neuron will not 
spike. This value plays a role similar to the voltage threshold 
for an action potential in the integrate-and-fire neuron. The 
minimum value reached in the mean conductance waveform can thus 
be identified with the `mean', since it determines the distance 
to threshold, and the variance in its value across cycles as the 
`variance'. The maximum and minimum value of the mean conductance 
waveform reached during an oscillation cycle is plotted as a 
function of $\sigma_{IV}$ in Fig.$~$\ref{InhibCond}Ca. The 
minimum increased and the maximum decreased as a function of 
$\sigma_{IV}$. The variance in the conductance at the time its 
mean has a minimum increased with $\sigma_{IV}$ 
(Fig.$~$\ref{InhibCond}Cb). The rate of increase with 
$\sigma_{IV}$ was higher for smaller values of $a_{IV}$. For 
large $a_{IV}$, the `variance' was small and an increase in 
$\sigma_{IV}$ would increase the `mean'. The corresponding effect 
was a shift of the $f-I$ curve. Another consequence of having a 
small `variance' is that the neuron became entrained, leading to 
saturation of the firing rate. Increasing $\sigma_{IV}$, for 
small $a_{IV}$, both increased the `variance' and the `mean', 
hence there was both a shift as well as a change of gain in the 
$f-I$ curve. This analysis indicates that for time-varying 
conductances the `variance' corresponds to the variability of the 
conductance waveform across trials and that this variability 
could modulate the gain of the neuron's response.\\
\indent
\subsection{Attentional modulation of $f-I$ curves is maximal 
for gamma frequency oscillations}

The strength of the modulation by synchrony was determined by the 
extent to which it could alter a neuron's firing rate. This 
change was quantified in terms of the ratio of the firing rate 
for moderate synchrony ($\sigma_{IV}=4$ ms) over the firing rate 
for weak synchrony ($\sigma_{IV}=10$ ms) because it reflected the 
strength of multiplicative interactions. This quantity had a peak 
at approximately $40$ Hz (Figure \ref{ResonanceCurve}). The 
resonance peak was mainly determined by the time constant of 
inhibition, $\tau_{inh}$. When $\tau_{inh}$ was increased the 
peak shifted to the left, and when $\tau_{inh}$ was decreased, 
the peak shifted to the right (data not shown). We studied how 
robust the resonance was. When the driving current was increased 
from $5$ to $6\mu A/cm^2$ or when the number of pulses per cycle 
was increased from $a_{IV}=25$ to $250$, the resonance 
disappeared. This indicates that for the resonance to occur the 
neuron needs to be subthreshold and there should be sufficient 
variability in the inhibitory inputs. The effect was moderately 
robust against jitter in the spike volley times: We obtained 
resonance peaks for $CV_T$ values up to approximately $0.05$.\\
\indent
\subsection{Modeling the experimental data}

Three key papers \cite{McAdams1999,Reynolds2000,Fries2001} 
reported experimental results on attentional modulation of the 
firing rate and the coherence of V4 neurons in macaques (see 
Appendix). In these papers, a stimulus was presented either 
inside or outside the receptive field of a neuron, and the focus 
of attention of the animal was either directed away from or into the 
receptive field by appropriate cues. The neurons that were 
recorded from were often orientation selective. We show here 
that the effects of attention in these studies can be explained 
by changes in the synchrony of local interneuron networks. As 
mentioned before, the network activity was not explicitly 
simulated, but instead the effects of modulating the synchrony 
on the principal neuron were modeled by dynamically changing 
$\sigma_{IV}$ during the trial. \\
\indent
The effects of attention were modeled by changing the parameters 
that control the synchronous inhibitory drive. In the attended 
state, input synchrony was increased by reducing the value of 
$\sigma_{IV}$ when compared with the non-attended state.   The 
receptive field and orientation selectivity were modeled as a 
constant depolarizing current $I$ or by an excitatory Poisson 
process. We made the cell orientation-selective 
\cite{MillerFerster2000,Shapley2003} by changing the amount of 
current injected into the neuron as a function of the stimulus 
orientation 
$\psi$: \( I_A=I_O+A_O \exp(-\psi^2/2\sigma_{\psi}^2)\), where 
$I_O$ is the baseline current in the presence of a stimulus, 
$A_O$ is the strength of orientation selectivity, and 
$\sigma_{\psi}$ represents the degree of selectivity.\\
\indent
Our aim was to show that modulating inhibitory synchrony can 
account for the experimentally reported effects of attention. 
Similar results were obtained for different sets of parameters. 
The experimental results were from different cortical areas 
and different conditions, and are not in complete agreement, so 
one unique set of parameters is unlikely to account for all 
experiments.  The different regimes needed to model the data from 
the different laboratories may provide insight into the state of 
the cortex in the different conditions.

\subsubsection{Computer Experiment I: Figure 4 of McAdams \& 
Maunsell \cite{McAdams1999}}

The effects of attention on the orientation-tuning curve were 
modeled as a decrease in the spike-time dispersion from 
$\sigma_{IV}=8$ to $7$ ms. The firing rate in the non-preferred 
orientation increased minimally from $4.5$ Hz to $5.5$ Hz in the 
attended state, whereas the firing rate in the preferred 
orientation increased by $20\%$ from $14.2$ Hz to $17.1$ Hz 
(Fig.$~$\ref{PrevExp}Aa), compatible with the experimental data. 
The asymptote was subtracted from the orientation-tuning curves, 
and the resulting curve in the non-attended state was rescaled 
by a factor $1.2$. After these manipulations, the curves for the 
attended and non-attended states overlapped (Figure 
\ref{PrevExp}Aa, inset), indicating, as observed experimentally, 
that the mean and width of the Gaussian had not changed.  
Similar results (data not shown) were obtained by decreasing the 
spike-time dispersion from $\sigma_{IV}=3$ to $2$ ms.  The 
firing rate in the non-preferred orientation increased from 
$1.3$ Hz to $4.2$ Hz in the attended state, whereas the firing 
rate in the preferred orientation increased from $13.6$ Hz to 
$24.7$ Hz.

\subsubsection{Computer Experiment II: Figure 8 of McAdams \& 
Maunsell \cite{McAdams1999}} Next, the temporal dynamics of 
attentional modulation in this figure were obtained using 
$\sigma_{IV}=2$ ms for the attended state and $\sigma_{IV}=3$ ms 
for the non-attended state (Fig.$~$\ref{PrevExp}Ab). The 
background firing rate was $3$ Hz and was not modulated by 
attention. There was an $8$ Hz temporal modulation, with a peak 
firing rate at stimulus onset equal to $34.4$ Hz. The firing-rate 
ratio between attended and non-attended state increased from $1$ 
at stimulus onset to approximately $1.5$ later in the trial, 
compatible with the experimental data.

\subsubsection{Computer Experiment III: Figure 1 of 
Fries et al. \cite{Fries2001}}

In the experiments of Fries et al. \cite{Fries2001}, the spikes 
from a neuron on one electrode were used to calculate the 
spike-triggered average (STA) of the LFP on another electrode.  
In the model, the LFP was estimated from the subthreshold membrane 
potential of a neuron that received theta-frequency excitatory 
and a gamma-frequency inhibitory synaptic inputs 
(Fig.$~$\ref{PrevExp}B). 
The STA in the model was calculated based on 
the spike train from another neuron receiving the same synaptic 
inputs.  The coherence of the inhibition was varied between the 
attended and non-attended state ($\sigma_{IV}=5$~ms to $4$~ms; 
$a_{IV}=5$ to $6$). There were no changes to the excitatory 
synaptic drive received by the neuron. However, the mean level of 
depolarization was varied to keep the mean firing rate constant 
(as explained in the Appendix, this also helps in estimating the 
SFC). We also analyzed two frequency bands, $5<f<15Hz$ (theta) 
and $34<f<44 Hz$ (gamma). The SFC in the theta range was $21\%$ 
less in the attended compared to the non-attended state, whereas 
the SFC in the gamma-frequency range more than doubled ($114\%$ 
increase, Fig.$~$\ref{PrevExp}Be).  
Our model does not predict the exact dynamics of the LFP.
The 
mean, phase and amplitude of this model LFP was different from 
the extracellularly recorded signals. Specifically, the 
gamma oscillations visible during the attended condition
in Fig.$~$\ref{PrevExp}Ba were much less pronounced 
than in experiment (Figure 1B of Ref.\cite{Fries2001}).
The modeled STA looks different from measured STA for the same
reason. However, the SFC is a ratio between the power spectrum of the STA 
and that of the LFP and might therefore be less
sensitive to the differences between the modeled and experimental LFP. 
Indeed, the changes in the model SFC with attention are qualitatively
similar to those reported in Ref.\cite{Fries2001}. There were
quantitative differences: the increase in SFC with attention
was of the order of 10\% in experiment, 
whereas in the model it was an order
of magnitude larger.  These differences could possibly be resolved
by a more detailed model that incorporates the electrical behavior 
of the extracellular medium in order to estimate the LFP more
accurately.

\section{Discussion}

\subsection{Summary}
The firing rate and coherence of a neuron can be modulated by the 
degree of synchrony of the inhibitory input. In the model 
investigated here, the mean number of inputs on each cycle, 
$a_{IV}$, was a key parameter in determining how the neuron's 
response properties depended on synchrony. The $a_{IV}$ is 
proportional to the number of neurons in the presynaptic network 
that are activated and the synaptic reliability (the probability 
that a presynaptic action potential results in a postsynaptic 
potential). The network activation is controlled by 
neuromodulators, such as acetylcholine, or by activation of 
metabotropic glutamate receptors \cite{GainSFN02}. It depends 
also on the extent of the chemical and gap-junction couplings 
within local networks \cite{Amitai2002} and, possibly, on longer 
range connections between different local networks 
\cite{Katz2003a,Katz2003b}. So $a_{IV}$ may reflect a combination 
of physiological parameters.\\
\indent
{\em Input synchrony modulates firing rate.} The firing rate 
increased when input synchrony was increased by reducing 
$\sigma_{IV}$. When $\sigma_{IV}$ was modulated dynamically, the 
change in firing rate was immediate and robust across trials. 
This implies that temporal dynamics of firing rate modulation is 
determined by how fast a presynaptic network can be activated, 
and how fast 
network activation results in an increase in synchrony. 
Interneuron networks can modulate their synchrony in a few 
oscillation cycles -- on the order of $100~ms$ for 
gamma-frequency range activity \cite{RapidMod} -- and the firing 
rate response is able to follow these rapid synchrony 
modulations. Rapid modulation of the firing rate observed in the 
cortex can therefore be due to the temporal dynamics of the 
stimulus as well as synchrony.  In addition to serving as a way 
to amplify the significance of information represented in a 
neural population, inhibitory synchrony provides an alternative 
pathway for cortical information transmission that can operate in 
parallel with changes in activity.  Our investigation also 
revealed that other statistical properties of the input, besides 
$\sigma_{IV}$, such as the oscillation frequency and $CV_T$, 
could also dynamically modulate the output firing rate (data not 
shown).\\
\indent
{\em Input synchrony modulates the sensitivity and gain of $f-I$ 
curves.} The $f-I$ curves were characterized in terms of 
sensitivity,  the weakest inputs to which the neuron will 
respond, and the gain, the rate of change of the firing rate 
with input amplitude. By increasing input synchrony, the 
neuron responded to weaker stimuli, and its gain increased so 
that the same increase in stimulus strength would result in a 
stronger increase of the firing rate. The relative size of the 
shift in sensitivity compared with the change in gain depended 
on $a_{IV}$. For small $a_{IV}\sim 10$, the change in gain 
dominated, whereas for larger $a_{IV}\sim 50$, the shift in 
sensitivity dominated. In the latter case, saturation was 
observed: when the firing rate was approximately equal to the 
oscillation frequency, it did not increase further when either 
the input was made stronger or when the input synchrony was 
increased. It should be noted that the unitary strength of the 
inhibitory inputs was normalized such that the mean (time-
averaged) inhibitory conductance remained constant when $a_{IV}$ 
was varied. This allowed us to distinguish the effects of an 
increase in mean inhibitory conductance with $a_{IV}$ from the 
effects of the reduced variability in the input with $a_{IV}$.\\
\indent
{\em Input synchrony modulates spike coherence with the network 
oscillation.} We measured the vector strength (VS) of the output 
spike train with respect to the oscillatory activity of the 
inhibitory input. Changes in the VS reflected the behavior of 
the SFC in the gamma-frequency range, but VS was easier to 
calculate and more robust than the SFC. The VS increased with 
input synchrony and it even did so when the firing rate remained 
the same.\\
\indent
{\em Input synchrony modulates the variability of spike trains 
within and across trials.} We determined the Fano factor (FF), 
representing the variability of the spike count across trials and 
the coefficient of variation (CV) -- the variability of 
interspike intervals during a trial. Generally, FF and CV 
decreased with increasing input synchrony. We found non-monotonic 
behavior of FF and CV as a function of the firing rate for 
$a_{IV}\sim 50$. Furthermore, the FF was sensitive to the jitter 
across trials in the phase of the oscillation at the start of the 
trial (or stimulus onset), whereas the CV was not sensitive to 
this phase.\\
\indent
{\em Firing rate modulation with input synchrony was most 
prominent for gamma-frequency oscillation.} The increase in 
firing rate induced by changing $\sigma_{IV}$ was maximal for 
$f_{osc}=40~Hz$ because of the time-constant of the inhibitory 
synapses. Networks of interneurons synchronize in the same 
frequency range \cite{traub,Buhl98}, indicating that the 
synchrony of their activity is well suited to modulate the firing 
rate of the pyramidal cells in cortex.

\subsection{Can attention modulate the synchrony of interneuron networks?}

Although the synchronization dynamics of inhibitory networks has 
been studied extensively using model simulations 
\cite{whittington,wang,white,CNS,MutInhibPub,TiesHippo,Soltesz2002b,Bartos2002}, the focus has almost exclusively been 
on the stationary state, rather than dynamic changes in 
synchrony. We found two types of networks whose synchrony can be 
changed by neuromodulators or excitatory neurotransmitters 
\cite{GainSFN02}. These results will be presented elsewhere; we 
summarize them here briefly. First, in a purely inhibitory 
network, synchrony can be modulated by increasing excitation to 
a part of the network. The activated neurons increase their 
firing rate and synchronize and reduce the activity of the other 
group of interneurons \cite{RapidMod}. Hence, the mean activity 
of the network that projects to a postsynaptic neuron, like the 
one studied here, remains approximately constant.  Synchrony can 
be modulated using this mechanism on time scales as short as 
$100~ms$. Second, in a mixed excitatory and inhibitory network, 
synchrony can be modulated by activating the interneuron network 
when the inhibitory and excitatory neurons are mode-locked to 
each other. In that case \cite{TiesHippo,Borgers2003,Brunel03}, 
synchronized excitatory activity recruits inhibitory activity 
that temporarily shuts down the excitatory activity. When the 
inhibition decays the excitatory neurons become active again and 
the cycle starts over. Activation of interneuron networks by 
neuromodulators may increase their synchrony, in turn increasing 
excitatory synchrony, but without altering the mean firing rate 
of individual neurons \cite{GainSFN02}.

\subsection{Gain modulation $f-I$ curves}

The statistics of the synaptic inputs determine the sensitivity 
and gain of the $f-I$ curve. Three mechanisms have been proposed 
for how multiplicative gain changes can be achieved 
\cite{Holt97,Emilio,TiesPRE,Doiron2001,Burkitt2001a,ChanceGain,Prescott2003,Silver2003,JM_GAIN,MillerGain2003,UlrichGain2003}. 
We briefly summarize them and discuss how they relate to gain 
modulation by inhibitory synchrony. \\
\indent
The response properties of neurons are different when they are 
driven by a supra- or infra-threshold currents \cite{TiesPRE}. 
In the former case neurons are tonically active and fluctuations 
in the input will not alter the firing rate, whereas in the 
latter case action potentials are induced by fluctuations 
(referred to as a fluctuation-dominated state \cite{TiesPRE}). 
The firing rates for neurons in the fluctuation-dominated state 
can be increased by either reducing the distance of the mean 
membrane potential to threshold, or by increasing the amplitude 
of the voltage fluctuations. For each of the mechanisms 
discussed below the neuron operates in the fluctuation regime, 
but the way that fluctuations increase or decrease the distance 
to threshold is different.\\
\indent
{\em Gain modulation by balanced synaptic inputs.} Under {\em in 
vivo} conditions neurons receive a constant barrage of 
excitatory and inhibitory inputs \cite{Shadlen98}. The synaptic 
inputs are called balanced when the effective reversal potential 
of the sum of excitatory and inhibitory inputs is equal to the 
neuron's resting membrane potential (leak reversal potential). 
By proportionally scaling the rates of excitatory and inhibitory 
inputs the amplitude of the voltage fluctuations can be 
modulated while maintaining a constant mean membrane potential. 
In the balanced mode the neuron is driven by fluctuations: the 
larger the fluctuations, the higher the firing rate. Chance and 
coworkers \cite{ChanceGain} found multiplicative gain modulation 
of the $f-I$ curves of neurons recorded in vitro experiments. 
Interestingly, an increase in balanced activity decreased the 
gain \cite{TiesPRE,Burkitt2001a,Burkitt2003}. The reason for 
this somewhat counterintuitive result is that the increase in 
input conductance dominates the increase in conductance 
variance, resulting in an amplitude reduction of voltage 
fluctuations. The saturation of dendritic nonlinearities can 
further enhance the change in gain obtained with balanced inputs 
\cite{Prescott2003}.  In a modeling study in which the 
excitatory and inhibitory fluctuations were independently 
varied, varying the amplitude of the inhibitory conductances was 
more effective than varying the amplitude of the excitatory 
conductances \cite{JM_GAIN}.\\
\indent
{\em Gain modulation by tonic inhibition and excitation.} Tonic 
inhibition by itself did not lead to multiplicative gain 
modulation \cite{Holt97,Doiron2001}. However, when tonic 
inhibition was applied in combination with either excitatory or 
inhibitory Poisson spike train inputs, changes in gain as well 
as shifts in sensitivity were observed 
\cite{Silver2003,UlrichGain2003}. Recently Murphy and Miller 
\cite{MillerGain2003} showed that changes in tonic excitation 
and inhibition can lead to approximate multiplicative gain 
modulation of cortical responses when the nonlinearity of the 
thalamic contrast response is taken into account.\\
\indent
{\em Gain modulation by correlations.} When a neuron is in a 
fluctuation-dominated state and receives inputs from different 
neurons, it is sensitive to correlations between these neurons. 
Stronger correlations lead to an increase in the amplitude of 
voltage fluctuations, hence to an increase in firing rate 
\cite{Emilio,Emilio2002}, because the mean input conductance is 
not altered by correlations, as was the case for balanced 
synaptic inputs.\\
\indent
{\em Gain modulation by inhibitory synchrony.}  Changing input 
synchrony for small $a_{IV}$ value resulted in a gain change of 
the $f-I$ curve. This mechanism is complementary to gain 
modulation by correlation. There are two different aspects of 
synchrony, the degree of coincidence -- how many neurons fire at 
approximately the same time, $a_{IV}$, and the precision  -- the 
temporal dispersion of the neurons that fire together, 
$\sigma_{IV}$. The number of neurons $a_{IV}$ roughly corresponds 
to the degree of correlation. Changing $\sigma_{IV}$ resulted 
both in a change of the distance to threshold as well as the 
amplitude of fluctuations. However, the `threshold' here is a 
conductance threshold: The neuron spiked when the inhibitory 
conductance became smaller than a threshold value. The mean 
minimum inhibitory conductance distance from the threshold and 
the fluctuations about this value determined the firing rate 
change. We are not aware of any mechanisms that have been proposed to 
selectively change the level of balanced input or degree of 
correlations in the network. In contrast, mechanisms 
have been proposed to modulate the input synchrony of the network, 
the parameter $\sigma_{IV}$ \cite{RapidMod}.\\
\indent
\subsection{Relation between attention and modulation of 
inhibitory synchrony}

More than a decade ago Crick and Koch proposed a link between 
oscillatory synchrony and attentional processing \cite{Crick90}. 
This led to a model in which excitatory neurons representing 
stimuli in the focus of attention produced correlated spike 
trains \cite{Niebur93}.  In their model, interneurons in cortical 
area V4 were activated by correlated spike trains and in turn 
suppressed V4 neurons responsive to stimuli outside the focus of 
attention; hence, the synchronized interneurons suppressed rather 
than enhanced activity. In the mechanism explored here, 
interneurons are active irrespective of the attentional state, 
and their degree of synchrony modulates the responsiveness of V4 
output neurons. The time-course of attentional modulation of 
neural responses was also studied by Deco and coworkers 
\cite{DecoTimeCourse2002,DecofMRI2002} but the synchrony of 
interneuron networks was not taken into account. \\
\indent
Under the assumption that attention acts by increasing the 
synchrony of interneuron networks, our model predicts that: (1) 
attention increases the coherence of spike trains with the local 
field potential; (2) the firing rate can increase with attention 
or remain the same depending on the stimulus strength; (3) 
attention can lead to a multiplicative gain change of firing 
rate response curves or to a shift in the sensitivity, depending 
on the extent of interneuron network activation by attention and 
the stimulus. Thus, changes in interneuron synchrony could 
potentially underlie a variety of seemingly unrelated 
observations. The size of the firing rate modulation predicted 
by this model agrees quantitatively with experimental 
observations (Section 3.5 and Appendix).  However, these results 
do not provide direct evidence that modulation of inhibitory 
synchrony is, in fact, responsible for the observed attentional 
effects. In the following we discuss specific predictions that 
derive from this hypothesis.\\ 
\indent 
McAdams \& Maunsell \cite{McAdams1999} observed multiplicative 
gain modulation of the orientation tuning curves of V4 neurons. 
We could reproduce these results quantitatively by changing 
$\sigma_{IV}$. The values of $\sigma_{IV}$ corresponding to the 
attended and non-attended state were not unique and we could 
obtain the same results with different combinations. The only 
constraint was that $\sigma_{IV}$ in the attended state had to 
be lower than in the non-attended state.  Multiplicative gain 
modulation of $f-I$ curves could simply be the consequence of a 
power law between $f$ and $I$ \cite{Hansel2002,Miller2002} and 
account for the contrast independence of orientation selectivity 
\cite{Anderson2000}. We could not fit our $f-I$ curves with a 
power-law and changing $\sigma_{IV}$ both changed the gain and 
shifted the sensitivity -- the gain change was not purely 
multiplicative. We could account for the results by McAdams \& 
Maunsell because the firing rate in response to a non-preferred 
stimulus -- the so-called asymptote -- was modulated by attention 
even though the background 
activity -- the firing rate without stimulus -- was 
not affected.
Only the modulation of the firing rate tuning curve 
above this asymptote was multiplicative. Hence, both a shift in 
sensitivity and a change in gain were required. The change in 
gain in the model only needed to be multiplicative over a 
limited range of the input current $I$. \\
\indent
Recent experiments were performed to determine whether attention 
would modulate the gain of the firing rate response, or whether 
the response properties could be interpreted as a shift in 
sensitivity \cite{Reynolds2000,Treue2002,Reynolds2003}. The 
firing rate elicited in response to a visual stimulus was 
measured as a function of stimulus contrast. The firing rate 
versus contrast had a sigmoidal shape and saturated. Gain 
modulation would imply that the saturation rate was increased by 
attention. This was not observed in the experiments and the 
changes were more consistent with a shift in sensitivity. Hence, 
depending on the specifics of the experimental protocol, 
attention can either increase sensitivity 
\cite{Reynolds2000,Treue2002,Reynolds2003} or gain 
\cite{McAdams1999,Treue1999}. Our model predicts that this 
behavior is correlated with the connectivity and degree of 
activation of the interneuron network.\\
\indent
McAdams \& Maunsell also reported that the attentional modulation 
of the firing rate reached its stationary value $500~ms$ after 
the onset of the response to the stimulus (Appendix). We could 
reproduce this by assuming that the $\sigma_{IV}$ value changed 
gradually from its background value (here equal to the 
$\sigma_{IV}$ in the non-attended state) to its value in the 
attended state. There was no change in $\sigma_{IV}$ if there was 
no stimulus. The model used here does not make predictions about 
the temporal dynamics of $\sigma_{IV}$ since that is a network 
property. However, if the change in firing rate is the result of 
modulating the inhibitory synchrony it implies that the stimulus 
needs to activate the interneuron network either directly in a 
bottom up fashion or indirectly through top down inputs.\\
\indent
Fries et al. \cite{Fries2001} reported that attention 
induced an increase in the gamma-frequency range coherence of the neuron's spike 
train with the LFP which was accompanied by only small changes in its firing 
rate.  We observed that a decrease in $\sigma_{IV}$ results in an 
increased coherence. This is expected based on general arguments. 
The amplitude of the inhibitory conductance waveform, defined as 
the distance between maximum and minimum value, increased with 
input synchrony (decreasing $\sigma_{IV}$). The spike timing 
precision of a neuron, which is directly related to the VS and 
$\sigma_{\phi}$, increased with the amplitude of the periodic 
driving current \cite{TiesQuasiPer}. The same holds for periodic 
and aperiodic stimulus waveforms {\em in vivo} 
\cite{Meister97,Reich97}. The small change in firing rate is 
indicative of saturation. We observed saturation for large 
$a_{IV}\sim 50$ when the neuron's firing rate was close  to the 
oscillation frequency of the inhibitory drive. The saturation of 
the firing rate was associated with a reduction in response 
variability on a given trial (CV) and across trials (FF).\\
\indent
There are other potential explanations for saturation. It can be 
due to the activation of intrinsic currents or increases in input 
conductance associated with synaptic background activity. These 
saturation effects would, however, in general not cause the 
reduction of response variability predicted by our model. In 
those experiments where attention did modulate the firing rate 
response, no concomitant reduction of response variability was 
observed in recordings of V4 neurons \cite{McAdams1999b}.

\subsection{Open problems and future work}

The attentional modulation of the firing rate of V4 neurons has 
been studied with two stimuli \cite{Luck1997,Reynolds1999},  a 
preferred stimulus that elicited a vigorous and robust response 
and a non-preferred stimulus that elicited a weaker response. 
When both stimuli were presented simultaneously in the neuron's 
receptive field, the firing rate was intermediate between the 
responses to each of the stimuli presented separately, rather 
than the sum of the two responses as a linear model would 
predict. This result was explained by a simple model proposed by 
Reynolds {\em et al} \cite{Reynolds1999}. Each stimulus 
activated a presynaptic population of neurons (in V2 in this 
case) that projected excitation and inhibition to the V4 neuron. 
The inhibitory component of the projection for the non-preferred 
stimulus was stronger and reduced the response to the preferred 
stimulus when both stimuli were presented simultaneously, 
consistent with the experimental observation. Their hypothesis 
was that attention would enhance synaptic efficacy of the 
presynaptic population of neurons representing the stimulus in 
the focus of attention, without increasing the activity of the 
presynaptic population. In their model, attention shifts the 
response toward the one that would be expected when the attended 
stimulus was presented alone, as indeed is observed 
experimentally.\\
\indent
In our model, the attentional enhancement in synaptic efficacy 
corresponds to an increase in synchrony of the inhibitory 
projection. This correctly predicts an increase in firing rate 
when attention is directed toward the preferred stimulus compared 
with attention directed outside the receptive field. However, 
when attention was directed toward the non-preferred stimulus, 
our model would still predict an increase in firing rate, albeit 
smaller, rather than the decrease observed in experiment. The 
underlying assumption was that attention operates bottom up from 
V2 to V4. An alternative hypothesis is that attentional 
modulation of synchrony is top down and operates on interneuron 
networks in V4 itself. This would predict that input synchrony 
increases when preferred stimuli in the receptive field are 
attended, but that synchrony decreases when the non-preferred 
stimuli are attended.\\
\indent
Neurons {\em in vivo} receive massive amounts of excitatory and 
inhibitory inputs \cite{Destexhe99}. Here we assumed that part of 
the inhibitory input was temporally modulated and observed that 
the firing rate could saturate when at the inhibitory oscillation 
frequency. The saturation rates observed during experiments may 
not correspond to those predicted by the model. It is likely that 
part of the excitation is also temporally modulated, allowing for 
the possibility that the neuron could saturate at rates 
determined by the time scale of temporal patterning of excitatory 
inputs. These inputs do not need to be oscillatory. Furthermore, 
in our model neuron there were no adaptation currents and there 
also was no synaptic coupling between neurons. An important issue 
for future study is how temporally modulated excitatory and 
inhibitory inputs derived from a network interact with intrinsic 
time-scales of the neuron (such as adaptation currents) to 
determine the saturation rate.


\newpage
\begin{table}[h]
{\tiny
\begin{tabular}{|l|l|l|l|l|l|l|l|}
\hline \hline
$\sigma_{IV}$&  $f$ (Hz) & $CV$ & FF & $\sigma_\phi$ & $VS$  & $SFC(\theta)$ & 
$SFC(\gamma)$\\
\hline \hline
Fig.$~$\ref{ExampleSpikeTrainRegular}  & & &  & & & & \\
\hline
$8~ms$&  4.40 (0.67)& 0.961 (0.137)  & 1.204 (0.189)& 0.189 (0.029) & 0.710 
(0.045) & 0.30 (0.58) & 0.14 (0.14)\\
$2~ms$&  18.26 (0.43)& 0.825 (0.031)  & 0.666 (0.086)& 0.096 (0.007) & 0.878 
(0.006) & 0.005 (0.002) & 0.026 (0.010)\\
\hline \hline
Fig.$~$\ref{ExampleSpikeTrain} &  & &  & & & & \\
\hline
$4~ms$&  22.33 (0.44) & 0.985 (0.038)  & 1.054 (0.327)& 0.181 (0.009)  & 0.685 
(0.012) & 0.006 (0.001) & 0.025 (0.015)\\
$2~ms$&  34.65 (0.49) & 0.781 (0.022)  & 0.646 (0.158)& 0.148 (0.007)  & 0.744 
(0.004) & 0.002 (0.001) & 0.038 (0.022)\\
\hline\hline
\end{tabular}}
\caption{ Spiking statistics for Figure \ref{ExampleSpikeTrainRegular} and 
\ref{ExampleSpikeTrain}.
The firing rate $f$, coefficient of variation $CV$, Fano factor $FF$, 
phase standard deviation $\sigma_\phi$, vector strength $VS$,
SFC in theta range (4.5-15 Hz) and gamma range (34-44 Hz) were calculated 
as described in Methods. Errors are given between parentheses and are 
the standard deviation across $10$ sets.}
\end{table}

\section{Appendix}
\subsection{Review of {\em in vivo} experimental results}

The roman numerals correspond to the subsections in Section 3.5.

{\em Experiment I.} McAdams \& Maunsell \cite{McAdams1999} 
recorded the response of a V4 neuron to Gabor patches (a 
sinusoidal grating multiplied by a 2-dimensional Gaussian 
density) that were presented for $500$ ms in its receptive field. 
The spatial frequency, color and size of the Gabor patch were 
chosen to elicit maximal responses. The contrast of the patch 
varied sinusoidally with a frequency of $4$ Hz. During the 
experiment the orientation of the Gabor patch was varied 
systematically. The responses of the neuron were recorded when 
the animal had to focus attention into the receptive field 
("attended") and when it had to focus onto a different Gabor 
patch at equal eccentricity away from the receptive field ("non 
attended").\\ \indent The mean firing rate of about $75\%$ of the 
cells that responded to this stimulus, could be fitted by a 
Gaussian-shaped orientation tuning curve in both the attended and 
non-attended states. There were four fitting parameters, the 
mean, width and amplitude of the Gaussian density and the 
asymptote (the firing rate in response to the least preferred 
stimulus orientation). We focused on Figure 4 in 
\cite{McAdams1999}, which showed population-averaged orientation-
tuning curves. The asymptote was approximately $5$ Hz in the 
non-attended state and increased slightly in the attended state. The 
amplitude of the Gaussian increased by $22\%$, going to $15$ Hz 
in the attended state  from $12$ Hz in the non-attended state. 
The mean and width of the Gaussian did not change significantly 
with attention.

{\em Experiment II.} Figure 8 of \cite{McAdams1999} shows the 
temporal dynamics of attentional modulation averaged across all 
responsive neurons. The background firing rate, before and after 
stimulus presentation, was $3.6$ Hz. It did not vary 
significantly with attention. There was an $8$ Hz stimulus-locked 
modulation in the firing rate with a peak firing rate of $35$ Hz 
at stimulus onset. The ratio of stimulus-induced firing in the 
attended state to that in the unattended state increased from 
unity at  stimulus onset to about $1.5$, $500$ ms after stimulus 
onset. The time course of this ratio was similar across different 
stimulus orientations.

{\em Experiment III.} A similar attentional paradigm was used in 
the experiments by Fries {\em et al} \cite{Fries2001}. They 
presented a pure luminance sinusoidal grating at $100\%$ contrast 
and optimal orientation in the receptive field of a V4 neuron for 
a random interval that lasted between $500$ ms and $5000$ ms. 
They recorded neural activity using $4$ electrodes that were 
spaced $650$ or $900\mu m$ apart. Multi-unit activity (elicited 
by the stimulus in the receptive field) was recorded on one 
electrode and the local field potential was recorded on a 
different electrode. The spike-triggered average (STA) on the LFP 
was calculated during stimulus presentation when the animal 
focused attention into the receptive field and when attention was 
directed away from the receptive field. The first $300$ ms after 
stimulus onset were discarded prior to their analysis. Changes of 
coherence with attentional state were assessed using the power 
spectrum density (PSD) of the STA, and the spike field coherence 
(SFC), which is the PSD of the STA normalized by the PSD of the 
LFP. The spectrum was divided into two frequency bands: $f<10 Hz$ 
(low frequency or theta) and $35<f<60 Hz$ (high frequency or 
gamma). The SFC in the low frequency range decreased by $23\%$ 
going from the non-attended to attended state, whereas the SFC in 
the gamma frequency range increased by $19\%$. The mean spike 
rate in the multi-unit recording did not change by more than 15\% 
with attentional state.

\subsection{Neuron model}
The equation for the membrane potential of the neuron was
\begin{equation}
C_m \frac{dV}{dt}=-I_{Na}-I_K-I_L-I_{syn}+I+C_m\xi,
\label{SINGNEUR}
\end{equation}
with the leak current
\(
I_L=g_L (V-E_L),
\)
the sodium current
\(
I_{Na}=g_{Na} m_{\infty}^3 h (V-E_{Na}),
\)
the potassium current:
\(
I_K=g_K n^4 (V-E_{K}),
\) and the synaptic current
$I_{syn}$ as described in the Methods section.
The intrinsic noise $\xi$ had zero mean and variance $2D$,
and $I$ was the injected current.
The channel kinetics were given in terms of $m$, $n$, and $h$.
They satisfied the following first-order kinetic equations,
\begin{equation}
\frac{dx}{dt}=\zeta (\alpha_x (1-x)-\beta_x x).
\end{equation}
Here $x$ labels the different kinetic variables $m$, $n$, and $h$, and
$\zeta=5$ was a dimensionless time-scale that
was used to tune the temperature-dependent
speed with which the  channels opened
or closed. The rate constants
were \cite{wang},
\begin{eqnarray}
\alpha_m&=& \frac{-0.1(V+35)}{\exp(-0.1(V+35))-1}, \nonumber\\
\beta_m&=& 4\exp(-(V+60)/18), \nonumber\\
\alpha_h&=&0.07\exp(-(V+58)/20), \nonumber\\
\beta_h&=&\frac{1}{\exp(-0.1(V+28))+1}, \nonumber\\
\alpha_n&=&\frac{-0.01 (V+34)}{\exp(-0.1(V+34))-1}, \nonumber\\
\beta_n&=&0.125~\exp(-(V+44)/80). \nonumber
\end{eqnarray}
We made the approximation that $m$ took the asymptotic value
$m_{\infty}(V(t)))=\alpha_m/(\alpha_m+\beta_m)$ instantaneously.
The standard set of values for the conductances used in this paper was
$g_{Na}=35$, $g_K=9$, and $g_L=0.1$ (in $mS/cm^2$),
and we took $E_{Na}=55~mV$, $E_K=-90~ mV$, and $E_L=-65~ mV$. The membrane
capacitance was $C_m=1\mu F/cm^2$.
$\Delta g_{inh}$ has units $\mbox{mS}/\mbox{cm}^2$;
$I$ is in $\mu \mbox{A}/\mbox{cm}^2$; $D$ is $mV^2/ms$; $f_{osc}$ is in Hz;
$P$ and $\sigma_{IV}$ are in ms; and $a_{IV}$ is dimensionless.

\subsection{Relation between the Spike Field Coherence and the Vector Strength}

The spike-triggered average (STA) of the LFP is the average LFP 
waveform around an output spike. The SFC is the ratio of the 
power spectrum density (PSD) of the STA and that of the LFP (see 
Methods). To obtain some intuition of the meaning of the SFC  
consider a simple example. Let the LFP be a 40 Hz cosine, the PSD 
then has a single peak at 40Hz.  Suppose neuron 1 produces action 
potentials on some cycles of the LFP at a fixed phase $\phi$. The 
STA is a 40 Hz cosine but shifted over $\phi$, yielding a PSD 
with a peak at 40 Hz (the PSD is not sensitive to the phase of 
the cosine). The SFC at 40 Hz is one and zero elsewhere. Suppose 
that the spike train of neuron 2 forms a homogeneous Poisson 
process with the same mean firing rate as neuron 1. The spikes 
are uncorrelated with the LFP and uniformly distributed in time. 
The STA would be equal to the mean of the LFP, zero in this case, 
and the PSD at 40 Hz would be zero, yielding an SFC equal to 
zero.\\
\indent
When there is jitter in the phase, with standard deviation 
$\sigma_\phi$, the SFC is $\exp(-4\pi^2\sigma_{\phi}^2)$. For 
this situation the SFC is completely determined by the phase 
jitter. The arrival times of the volleys of inhibitory inputs are 
known in the model simulations, hence the phase of spike times 
with respect to the oscillation are known. The phase is a 
cyclical variable, which means that $\sigma_{\phi}$ is not a good 
measure of the phase jitter (it depends on the mean phase and the 
jitter), but the vector strength is,  $VS=\vert \langle 
\exp(i2\pi\phi) \rangle\vert = \sqrt{\langle \cos(2\pi \phi) 
\rangle^2 +\langle \sin(2\pi \phi)\rangle^2}$. For our example, 
we obtain $VS=\exp(-4\pi^2\sigma_{\phi}^2)$=SFC, such that $VS=1$ 
for perfect phase-locking (Neuron 1) and VS is zero for a uniform 
phase distribution such as would be the case for Neuron 2.  Our 
analysis yields a potentially important insight into the 
experimental results of Fries et al. \cite{Fries2001}. When the 
amplitude of the LFP is increased, but the spike times stay 
exactly the same, the amplitude of the STA increases 
proportionally and the SFC stays the same. However, when the 
phase jitter of the spike times decreases at the same time, the 
SFC will increase. Previously we found that for noisy neurons 
driven by sinusoidal current the jitter decreases when the 
amplitude of the current increases \cite{ReliSFN}.  This follows 
because the spike-time jitter is proportional to $dV/dt$ at the 
action potential threshold \cite{Chialvo00}, and because $dV/dt$ 
is proportional to the stimulus amplitude. The LFP is in general 
not a cosine but contains other frequency components, hence, the 
VS is not equal to the SFC under general conditions, but changes 
in the VS faithfully represent changes in the coherence in the 
gamma-frequency range which is the subject of this study.\\
\indent
These theoretical results for the STA are based on averages over 
the full distribution of spike times. In computational 
experiments the STA is sampled using a finite number of spike 
times. For $\sigma_{\phi}=0$, the STA obtained using only one 
spike is equal to its theoretical value. However, for neuron 2, 
the STA for one spike would be a cosine, and the SFC would be 
one. The theoretical STA, equal to zero, can only be obtained by 
adding many cosines with random phases. This explains the 
counterintuitive results for the SFC calculated based on 
Fig.$~$\ref{ExampleSpikeTrainRegular}: The firing rate was so 
low, that the STA was not correctly sampled for the given trial 
length. The calculation of the SFC is illustrated in 
Fig.$~$\ref{ExampleSpectral}.  

\subsection{Parameter values for the figures}

{\em Figure \ref{ExampleSpikeTrainRegular}} 
The neuron was driven by a synchronous inhibitory input with 
$a_{IV}=25$, $\Delta g_{inh}=0.044 mS/cm^2$, $\tau_{inh}=10 ms$, 
$P=26.10$ ms, $f_{osc}=38.3$ Hz, $CV_T=0.095$, $I=4.0 \mu 
A/cm^2$, $D=0.08 mV^2/ms$. For $t<300$ and $t>700$, the spike 
time dispersion $\sigma_{IV}$ was $8~ms$, and, for $300\le t \le 
700$, it was $\sigma_{IV}=2~ms$. The subthreshold membrane 
potential in (b) was obtained by reducing the injected current to 
$I=1.0 \mu A/cm^2$ to prevent action potentials. The firing rate 
histogram was calculated based on  $500$ trials.

{\em Figure \ref{ExampleSpikeTrain}} 
The neuron was driven by a synchronous inhibitory input with 
$a_{IV}=10$, $\Delta g_{inh}=0.11 mS/cm^2$, $\tau_{inh}=10 ms$, 
$P=26.10$ ms, $f_{osc}=38.3$ Hz, $CV_T=0.095$, $I=2.4 \mu 
A/cm^2$, $D=0.04 mV^2/ms$. For $t<1000$ and $t>2000$, the spike 
time dispersion $\sigma_{IV}$ was $4ms$, and, for $1000\le t \le 
2000$, it was $\sigma_{IV}=2ms$. There was also a temporally 
homogeneous excitatory input with rate $\lambda_{exc}=1000$ Hz, 
$\Delta g_{exc}=0.02 mS/cm^2$ and $\tau_{exc}=2 ms$. The 
subthreshold membrane potential in (b) was obtained by reducing 
current to $I=-2.0 \mu A/cm^2$ to prevent action potentials. The 
firing rate histogram was calculated based on $500$ trials.

{\em Figure \ref{FirSurf1}}
The neuron was driven by a synchronous inhibitory input with 
$a_{IV}=50$, $\Delta g_{inh}=0.022 mS/cm^2$, $\tau_{inh}=10 ms$, 
$P=26.10$ ms, $f_{osc}=38.3$ Hz, $CV_T=0.0$, $D=0.0 mV^2/ms$. 
Parameters were on a two-dimensional grid defined by: $I=2.0$ to 
$I=7.5~\mu A/cm^2$ in steps of $0.1 ~\mu A/cm^2$ and 
$\sigma_{IV}=1~ms$ to  $\sigma_{IV}=6~ms$ in steps of $0.5~ms$. 
For each parameter value, statistics were over $20$ trials of 
$3000~ms$ length.

{\em Figure \ref{FirSurf2}}
The parameters of the inhibitory and excitatory drive were the 
same for Fig.$~$\ref{ExampleSpikeTrain}, except that 
$\sigma_{IV}$ and $I$ were varied on a two-dimensional grid 
defined by: $I=0.0$ to $I=5.0~\mu A/cm^2$ in steps of $0.1 ~\mu 
A/cm^2$ and $\sigma_{IV}=1~ms$ to  $\sigma_{IV}=10~ms$ in steps 
of $1~ms$. For each parameter value, statistics were over $20$ 
trials of $3000~ms$ length.

{\em Figure \ref{fICurves}}
(A) Parameter values were $P=26.08$ ms, $f_{osc}=38.35$ Hz,  
$CV_T=0$, $a_{IV}=10$, $g_{inh}=0.11 mS/cm^2$, $\tau_{inh}=10$ 
ms, $D=0 mV^2/ms$. From left to right, $\sigma_{IV}=1$, $2$, $3$, 
$4$, and $5$ ms. Inset, data for $\sigma_{IV}=2$, $3$, $4$, and 
$5~ms$ was rescaled according to $f\rightarrow f/\lambda_f$ and 
$I\rightarrow (I-\Delta_I)$ to coalesce with the curve for 
$\sigma_{IV}=1$ ms.  (B) Parameter values were $P=26.08$ ms, 
$f_{osc}=38.35$ Hz,  $CV_T=0$, $a_{IV}=50$, $g_{inh}=0.022 
mS/cm^2$, $\tau_{inh}=10$ ms, $D=0 mV^2/ms$. From left to right, 
$\sigma_{IV}=1$, $3$ and $5$ ms.  A sigmoidal function, 
$f(I)=\frac{A}{2}(1+\tanh(\lambda_I (I-\Delta_I))$, was fitted to 
the $f-I$ curves. In (A-B) a transient of $100$ ms was discarded 
before analysis. The firing rate was then calculated over a $50$ s
time interval.

{\em Figure \ref{fICurvesExtra}}
Parameter values were $P=26.08$ ms, $f_{osc}=38.35$ Hz,  
$CV_T=0$, $\tau_{inh}=10$ ms, $\sigma_{IV}=1~ms$, and $D=0 mV^2/ms$.
In (A,C) $a_{IV}=10$, $g_{inh}=0.11 mS/cm^2$ and
(B,D) $a_{IV}=50$, $g_{inh}=0.022mS/cm^2$.
From left to right (A,B) or as indicated by the direction
of the arrow (C,D), the input rate was $100$, $300$, $500$, $700$
and $900$ EPSPs per second.  A transient of $50$ ms was discarded 
before analysis. The firing rate was then calculated over $20$ trials
of $3$ s duration. 

{\em Figure \ref{ResonanceCurve}}
Parameter values were $P=26.08$ ms, $f_{osc}=38.35$ Hz,  
$CV_T=0$, $a_{IV}=25$, $g_{inh}=0.044 mS/cm^2$, $\tau_{inh}=10$ 
ms, $I=5.0 \mu A/cm^2$, $D=0 mV^2/ms$. Mean firing rate was 
calculated over $50$s after discarding a $100$ms transient. The 
dynamical range was defined as the ratio of the firing rate for 
$\sigma_{IV}=4$ms divided by that for $\sigma_{IV}=10$ ms 
provided the latter is nonzero.

{\em Figure \ref{PrevExp}}
(Aa) Neurons received orientation-tuned input current, 
\(I=I_O+A_O \exp(-\psi^2/2\sigma_{\psi}^2)\) with  $I_O=5.5\mu 
A/cm^2$, $A_O=1.3\mu A/cm^2$, $\sigma_\psi=30^\circ$.  $\psi$ 
took values between $-90^\circ$ and $+90^\circ$ with 
discretization step $\Delta\psi=9/4^\circ$. The firing rate was 
obtained from pre-calculated $f-I$ curves using interpolation. 
Parameters were: $\sigma_{IV}=8$ ms for attention directed away 
and $\sigma_{IV}=7$ ms for attention directed into the receptive 
field; $P=26.08$ ms, $f_{osc}=38.35$ Hz, $CV_T=0$, $a_{inh}=25$, 
$\Delta g_{inh}=0.044 mS/cm^2$, $\tau_{inh}=10 ms$, $D=0 
mV^2/ms$. To mimic neuronal variability, normally distributed 
noise with a standard deviation equal to 5$\%$ of the firing rate 
was added to the firing rate. Solid curves are a 4-point running 
average, only every fourth point is shown in the graph.\\

(Ab) Neurons received a time-dependent driving current $I(t)$ and
an inhibitory synchronous drive with time-varying parameters
$a_{IV}(t)$. In the non-attended state $\sigma_{IV}=3$ ms and it
was $\sigma_{IV}=2$ ms in  the attended state.
For $t<500$,
the current was $I(t)=0$;
for $500 \le t < 1000$, it was
$I(t)=0.8+0.56 \left[\sin (2\pi t/250)\right]^2+ 2.4 \min\left[(t-
500)/350,1\right]$;
for $1000 \le t < 1100$, it was $I(t)=3.2\left[1-(t-1000)/100\right]$; and
for $t\ge 1100$, the current was zero.
For $t<500$, the inhibitory activity was $a_{IV}=5$;
for $500\le t < 850$, $a_{IV}=5+20\left[(t-500)/350\right]$;
for $850 \le t < 1000$, it was $a_{IV}=25$;
for $1000 \le t < 1100$, it was $a_{IV}=25-20\left[(t-1000)/1000\right]$; and
for $t\ge 1100$, it was $a_{IV}=5$.
$I$ was in $\mu A/cm^2$, $t$ and $\sigma_{IV}$ in ms.
Other parameters were $P=26.39$ ms, $f_{osc}=37.90$ Hz, $CV_T=0.11$, $D=0.4 
mV^2/ms$.\\

(B) The neuron received synchronous inhibitory and excitatory 
drives. Parameters for the excitatory drive were $P=116$ ms, 
$f_{osc}=8.62$ Hz, $CV_T=0$, $a_{EV}=25$, $\sigma_{EV}=30$ ms. 
For the inhibitory drive they were $P=26.08$ ms, $f_{osc}=38.35$ 
Hz, $CV_T=0$, with, for the non-attended state, (Ba) 
$\sigma_{IV}=5$ ms, $a_{IV}=5$, $I=1.6 \mu A /cm^2$; and for the 
attended state, (Bb) $\sigma_{IV}=4$ ms, $a_{IV}=6$, $I=1.8 
\mu A /cm^2$. In both cases, $D=0 mV^2/ms$. The local field 
potential was calculated by hyperpolarizing the neuron (the 
injected current was reduced to $I=0.0 \mu A/cm^2$). (Bc) Spike 
triggered average was calculated based on a 10 second long LFP 
waveform of which the first 200 ms was discarded as a transient. 
Sampling rate was 5 kHz (temporal resolution was $0.2$ ms). (Bd) 
Power spectrum density was calculated based on the STA sampled at 
4096 points at a temporal resolution of 0.2 ms, there were 
$n_{fft}=2048$ points in the Fourier transform.

\newpage
\begin{figure}
\caption{ 
Model for generating synchronous spike volleys in a simulated 
population of interneurons. (A) Spike volleys arrived at a rate 
of $f_{osc}$ volleys per second (the mean time separation between 
volleys was the period $P=1/f_{osc}$). (B) Spike-time probability 
was generated by convolving spike volley times with a Gaussian 
filter; its width was $\sigma_{IV}$ and its area (sum of the 
bins) was $a_{IV}$. (C) Spike times were generated as a Poisson 
process from the spike time probability. (D) Each input spike 
caused an exponentially decaying inhibitory conductance pulse 
with a unitary conductance $\Delta g_{inh}$ and a decay constant 
$\tau_{inh}$. (E) Concomitant voltage fluctuations in the 
postsynaptic neuron. } \label{Methods}
\end{figure}

\begin{figure}
\caption{
Inhibitory input synchrony gated neural activity. (A) The 
membrane potential, (B) the local field potential (LFP), and (C) 
the firing rate as a function of time. (D) Rastergram of the 
first ten trials. During the time interval between $t=300$ and 
$700$ ms (indicated by the bar in (D)), $\sigma_{IV}$ was reduced  
to $2$ ms from $8$ ms. The full parameter set is given in the 
Appendix.
}
\label{ExampleSpikeTrainRegular}
\end{figure}

\begin{figure}
\caption{
The inhibitory input synchrony modulated the output firing rate. 
(A) The membrane potential, (B) the local field potential (LFP), 
and (C) the firing rate as a function of time. (D) Rastergram of 
the first ten trials. During the time interval between $t=1000$ 
and $2000$ ms (indicated by the bar in (D)), $\sigma_{IV}$ was 
reduced to $2$ ms from $4$ ms. The full parameter set is given in 
the Appendix.
}
\label{ExampleSpikeTrain}
\end{figure}

\begin{figure}
\caption{The firing rate and the coherence for $a_{IV}=50$. (A) 
Firing rate $f$ as a function of $\sigma_{IV}$ and $I$. (B) 
Coefficient of variation, (C) Vector strength, and (D) Fano 
factor as a function of firing rate. The solid lines in (B,D) are 
3-pt running averages for, from bottom to top, $\sigma_{IV}=1$, 
$2$, $\cdots$, $5~\mbox{ms}$, circles are the data points. The 
same $\sigma_{IV}$ values are shown in (C), but now ordered from 
top to bottom.
}
\label{FirSurf1}
\end{figure}

\begin{figure}
\caption{The firing rate and the coherence for $a_{IV}=10$ together 
with a background excitatory synaptic input. Panels are as in
Fig.$~$\ref{FirSurf1}, parameters are as in Fig.$~$\ref{ExampleSpikeTrain}. 
For clarity, we show only the 3-point running
average in panel D.
}
\label{FirSurf2}
\end{figure}

\begin{figure}
\caption{Subtractive and divisive scaling of $f-I$ curves with 
inhibitory synchrony in the computer model. (A) Multiplicative 
gain modulation with inhibitory synchrony. $a_{IV}=10$, from top 
to bottom $\sigma_{IV}=1$, $2$, $3$, $4$ and $5$ ms. Inset: all 
curves could be collapsed by a shift in the current and a 
rescaling of the firing rate axis. (B) Shift in neural 
sensitivity with inhibitory synchrony, $a_{IV}=50$, from left to 
right, $\sigma_{IV}=1$, $3$, and $5$ ms. The solid lines are fits 
to a sigmoid function, filled circles are the simulation results. 
(C,D) Fitting parameters as a function of $\sigma_{IV}$. (C) The 
shift $\Delta_I$ (circles) and firing rate gain $\lambda_f$ 
(squares) necessary to make the curves in (A) collapse on the 
$\sigma_{IV}=1$ reference curve. (D) The midpoint $\Delta_I$ 
(circles) and slope $\lambda_I$ (squares) of the best-fitting 
sigmoid. The asterisk labels $f-I$ curves that were not well 
fitted by a sigmoid.
}
\label{fICurves}
\end{figure}
\begin{figure}
\caption{
The difference between $f$-$I$ curves for $a_{IV}=10$ and $50$ is robust
against excitatory Poisson spike train inputs. The (A,B) firing rate
and (C,D) coefficient of variation is plotted versus current for
(A,C) $a_{IV}=10$ and (B,D) $a_{IV}=50$. 
The jitter is $\sigma_{IV}=1~ms$. The input rate was $100$, $300$, $500$, $700$ and 
$900$ EPSPs per second increasing from right to left (A,B) or as indicated
by the direction of the arrow (C,D).
}
\label{fICurvesExtra}
\end{figure}
\begin{figure}
\caption{Variability of the inhibitory conductance depends on the 
synchrony $\sigma_{IV}$ and the mean number of inputs $a_{IV}$. 
(Aa) Inhibitory conductance as a function of time for $a_{IV}=10$ 
and $\sigma_{IV}=1 ~ms$, we plot the mean across trials (solid 
line) and a sample trace for one trial (dashed line). (Ab) The average over 
cycles (solid line) for one trial is the same as the average of 
one cycle across multiple trials. The dashed curves are the mean 
plus or minus twice the standard deviation across trials. (Ba) 
Increasing the mean number of inputs per cycle to $a_{IV}=100$ 
(thick solid line)
from $a_{IV}=10$ (dashed line) decreases the standard deviation 
in the conductance (arrows). (Bb) Decreasing the degree of input 
synchrony to $\sigma_{IV}=5~ms$ reduces the temporal modulation 
of the conductance. (Ca) The maximum (top) and minimum (bottom) 
of the mean across cycles of the conductance waveform as a 
function of $\sigma_{IV}$. The mean did not depend on the value 
of $a_{IV}$. (Cb) The standard deviation of the conductance 
values across cycles at the phase when the maximum (top) or 
minimum (bottom) value of the mean is reached. Data is for 
$a_{IV}=10$ (solid line) and $a_{IV}=100$ (dashed line).
}
\label{InhibCond}
\end{figure}

\begin{figure}
\caption{
Gamma-frequency-range resonance in the strength of attentional 
modulation of the firing rate. Strength of attentional modulation 
is quantified as the ratio of the firing rate for $\sigma_{IV}=4$ 
ms over that for $\sigma_{IV}=10$ ms. The ratio attains its 
maximal value at approximately $40$ Hz.
}
\label{ResonanceCurve}
\end{figure}

\begin{figure}
\caption{ 
Modulation of inhibitory synchrony {\em in model simulations} 
reproduced attentional modulation of V4 neurons observed in 
experiment. (A) Model of attentional modulation in McAdams $\&$ 
Maunsell \cite{McAdams1999}. The neuron received synchronous inhibitory 
input. (a) Firing rate as a function of stimulus orientation for 
two conditions: (solid lines, filled symbols) 
attention was directed away from the 
receptive field, $\sigma_{IV}=8$ ms and 
(dashed lines, open symbols) attention was 
directed into the receptive field, $\sigma_{IV}=7$ ms. Inset: The 
two curves coalesced when the asymptotic firing rate was 
subtracted and the residual of the solid line was rescaled by a 
factor $1.2$ along the $y$-axis. (b) Temporal dynamics of 
attentional modulation of the firing rate. The bar indicates the 
presence of a driving current representing the presence of a 
stimulus in the receptive field. (Solid line) Attention directed 
away from receptive field, $\sigma_{IV}=3$ ms and (dashed line) 
into the receptive field, $\sigma_{IV}=2$ ms. (B) Model of 
attentional modulation reported in Fries et al. \cite{Fries2001}. 
A neuron 
received synchronous inhibitory input in the gamma-frequency 
range, and excitatory input in the theta-frequency range. For 
the solid lines attention was directed away from the receptive 
field, $\sigma_{IV}=5$ ms, and for the dashed lines attention was 
directed into the receptive field, $\sigma_{IV}=4$ ms. (a-b, top) 
The local field potential (LFP) and (bottom) a spike train from 
one neuron. (c) The spike-triggered average (STA) of the LFP, the 
solid line was shifted by +2mV for clarity. (d) Power spectrum 
density (PSD) of the STA. (e) The spike field coherence (SFC). 
The full set of parameter values is given in the Appendix. } 
\label{PrevExp}
\end{figure}

\begin{figure}
\caption{Calculation of the Spike Field Coherence (SFC) for the model parameters
used for Fig.$~$\ref{ExampleSpikeTrainRegular}.
The model parameters are the same except that for $\sigma_{IV}=8~ms$
we took $I=6.0 \mu A/cm^2$.
(A) Spike triggered average (STA) of the LFP for (a) $\sigma_{IV}=2~ms$
and (b) $8~ms$.
(B) The power spectrum (PSD) of (a) the LFP and  (b) the
STA for (top)  $\sigma_{IV}=2~ms$ and (bottom) $8~ms$.
(C) SFC for (a) $\sigma_{IV}=2~ms$ and (b) $8~ms$.
}
\label{ExampleSpectral}
\end{figure}

\end{document}